\documentclass[prl,twocolumn,superscriptaddress,floats,showpacs]{revtex4}

\usepackage[dvips]{graphicx}
\usepackage{amsmath}
\usepackage{booktabs}
\usepackage{dcolumn}

\newcommand{\nbf}{Ba$_\text{1-x}$Na$_\text{x}$Fe$_\text{2}$As$_\text{2}$}
\newcommand{\bnf}{Ba$_\text{1-x}$Na$_\text{x}$Fe$_\text{2}$As$_\text{2}$}
\newcommand{\bnfdf}{Ba$_\text{0.65}$Na$_\text{0.35}$Fe$_\text{2}$As$_\text{2}$}

\newcommand{\bfa}{BaFe$_\text{2}$As$_\text{2}$}

\begin{document}

\advance\vsize by 2 cm

\title{Spin reorientation in Na-doped BaFe$_2$As$_2$ studied by neutron diffraction }

\author{F. Wa\ss er}
\affiliation{II. Physikalisches Institut, Universit\"{a}t zu
K\"{o}ln, Z\"{u}lpicher Stra\ss e 77, D-50937 K\"{o}ln, Germany}

\author{A. Schneidewind}
\affiliation{J\"ulich Centre for Neutron Science,
Forschungszentrum J\"ulich GmbH, Outstation at MLZ,
Lichtenbergstr. 1, 85748 Garching, Germany}

\author{Y. Sidis}
\affiliation{Laboratoire L\'eon Brillouin, C.E.A./C.N.R.S., F-91191 Gif-sur-Yvette CEDEX, France}

\author{S. Wurmehl}
\affiliation{%
Leibniz-Institut f\"ur Festk\"orper- und Werkstoffforschung
Dresden, Helmholtzstra\ss e 20, D-01069 Dresden, Germany}
\affiliation{Institut f\"ur Festk\"orperphysik, Technische
Universit\"at Dresden, D-01171 Dresden, Germany}
\author{S. Aswartham}
\affiliation{%
Leibniz-Institut f\"ur Festk\"orper- und Werkstoffforschung
Dresden, Helmholtzstra\ss e 20, D-01069 Dresden, Germany}
\author{B. B\"uchner}
\affiliation{%
Leibniz-Institut f\"ur Festk\"orper- und Werkstoffforschung
Dresden, Helmholtzstra\ss e 20, D-01069 Dresden, Germany}
\affiliation{Institut f\"ur Festk\"orperphysik, Technische
Universit\"at Dresden, D-01171 Dresden, Germany}

\author{M. Braden}
\email{braden@ph2.uni-koeln.de} \affiliation{II. Physikalisches
Institut, Universit\"{a}t zu K\"{o}ln, Z\"{u}lpicher Stra\ss e 77,
D-50937 K\"{o}ln, Germany}

\date{\today}


\begin{abstract}

We have studied the magnetic ordering in Na doped BaFe$_2$As$_2$
by unpolarized and polarized neutron diffraction using single
crystals. Unlike previously studied FeAs-based compounds that
magnetically order, Ba$_{1-x}$Na$_x$Fe$_2$As$_2$ exhibits two
successive magnetic transitions: For x=0.35 upon cooling magnetic
order occurs at $\sim$70\ K with in-plane magnetic moments being
arranged as in pure or Ni, Co and K-doped BaFe$_2$As$_2$ samples.
At a temperature of $\sim$46\ K a second phase transition occurs,
which the single-crystal neutron diffraction experiments can
unambiguously identify as a spin reorientation. At low
temperatures, the ordered magnetic moments in
Ba$_{0.65}$Na$_{0.35}$Fe$_2$As$_2$ point along the $c$ direction.
Magnetic correlations in these materials cannot be considered as
Ising like, and spin-orbit coupling must be included in a
quantitative theory.

\end{abstract}

\maketitle

There are two promising explanations for the appearance of
high-temperature superconductivity in FeAs-based materials
\cite{1}. Orbital fluctuations may result in a s$^{++}$
superconducting state \cite{2,3} and can reflect the fact that
highest superconducting transition temperatures arise in materials
with almost ideal FeAs$_4$ tetrahedrons \cite{4} and, thus, with
highest orbital degeneracy. On the other hand, there are strong
magnetic fluctuations associated with the antiferromagnetic (AFM)
order in the parent compounds which can explain a s$^\pm$
superconducting state \cite{6}.

\begin{figure}[t]
\begin{center}
\includegraphics*[width=.75\columnwidth]{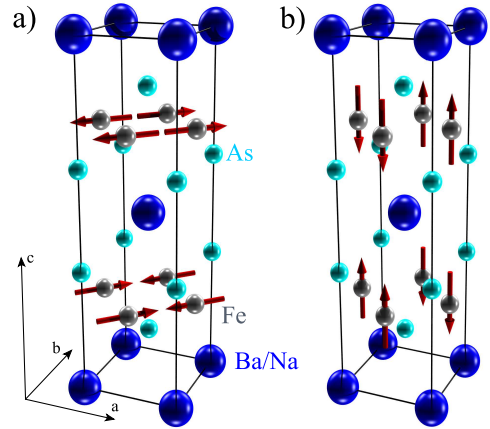}
\end{center}
\caption{(Color online) Magnetic structure of \nbf \ in the AFM1
phase $46 < $ T $< 70$\ K and in the AFM2 phase at low
temperature, T $< 46$\ K. Note that the fourfold axis is broken in
both cases by the the in-plane component of the magnetic
propagation vector chosen as (0.5,0.5,1). \label{fig1}}
\end{figure}

Magnetism and orbital degrees of freedom are closely tied in
FeAs-based compounds. Although the structural distortion
accompanying AFM order in the parent materials remains small
\cite{7,7b,8} its electronic signatures are rather strong as seen
in the anisotropic resistance \cite{rhoa,rhob}, in angle-resolved
photoemission studies (ARPES) \cite{9a,9b} or optical spectroscopy
\cite{10}. In addition the magnon dispersion in the AFM state is
fully anisotropic \cite{11} inspiring theoretical models of
orbital order driving magnetic interaction similar to those
applied to manganates \cite{orbo1,orbo2,orbo3}. Studying the
interplay between orbital and magnetic degrees of freedom seems
crucial for the understanding of FeAs-based materials.

Here we focus on spin-space anisotropies arising from the
spin-orbit coupling between spin and orbital moments and which
thus allow for a direct view on orbital contributions. In AFM
BaFe$_2$As$_2$, polarized  neutron scattering shows that it costs
more energy to rotate the magnetic moment within the planes than
perpendicular to them \cite{12}. Magnetic anisotropy clearly
persists into the superconducting range of the phase diagrams
\cite{13,13a,13b,13c,13d}.

So far all AFM ordered FeAs-based compounds exhibit a single
magnetic transition to a magnetic structure where moments are
aligned parallel to the in-plane component of the magnetic
propagation vector, which is (0.5,0.5,1) \cite{7,7b,8} (we use
reduced lattice units with respect to $\frac{2\pi}{a}$ and
$\frac{2\pi}{c}$ with $a\sim$3.9\ and $c\sim$13.1\ \AA ). Very
recently a second magnetic transition was reported for Na-doped
BaFe$_2$As$_2$ and interpreted as a microscopic superposition of
the two equivalent ordering schemes, with propagation vectors
(0.5,0.5,1) and (-0.5,0.5,1), into a two-$k$ structure \cite{14}.
We have studied single crystalline samples of \bnf \ at various
doping levels finding clear evidence for a reorientation of the
spins occurring at intermediate doping at low temperature. This
spin reorientation towards alignment of the magnetic moments
parallel to the c direction agrees with the anisotropies observed
in pure and in Co-doped \bfa .

Single crystals of \bnf \ were grown using a self-flux high
temperature solution growth technique. Details of sample growth
and characterization can be found in reference \cite{15}. Due to
the air-sensitivity of \bnf \ all synthetization procedures as
well as the mounting and orienting of the crystals into the sample
cans were done in an Ar box. Lattice parameters of the large
crystals were determined by neutron diffraction using the 3T1
diffractometer at the Orph\'ee reactor in Saclay ($k_i$=2.66\
\AA$^{-1}$). There are two reports on the phase diagram of \bnf \
in the literature \cite{16,17}, both based on polycrystalline
samples. The lattice constants and magnetic properties determined
with our single crystals fit to the results reported by Cortes-Gil
et al. \cite{17} while there seems to be an offset in the
Na-content when comparing with the data described by Avci et al.
\cite{16}. Neutron scattering experiments aiming at the
characterization of the magnetic structure as function of
temperature were performed at the PANDA triple-axis spectrometer
(Maier-Leibnitz Zentrum, Garching, $k_i$=1.55 and 2.57\
\AA$^{-1}$), at 3T1 and at the 4F1 triple-axis spectrometer (both
at the Orph\'ee reactor)). On the 4F1 spectrometer we used a
neutron beam with $k_i$=2.57\ \AA$^{-1}$ polarized by a bender and
analyzed the final polarization by a Heusler crystal. We performed
longitudinal polarization analysis by guiding the neutron spins
with a set of Helmholtz coils. In all experiments either
pyrolithic graphite or Be filters were used to suppress higher
order contaminations and crystals were oriented in the [110]/[001]
scattering planes.

\begin{figure}[t]
\begin{center}
\includegraphics*[width=.8\columnwidth]{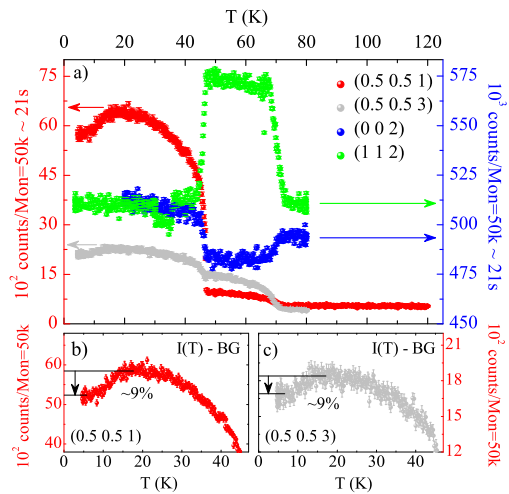}
\end{center}
\caption{(Color online) (a) Temperature dependence of magnetic and
nuclear Bragg reflection intensities in \bnfdf  \ measured on the
PANDA spectrometer. Note that the magnetic, (0.5,0.5,1) and
(0.5,0.5,3), and nuclear, (112) and (002), Bragg peaks refer to
different scales respectively. (b) and (c) show an enlargement of
the low-temperature range illustrating the reduction of magnetic
intensities in the superconducting state. \label{fig2}}
\end{figure}

Fig. 1 shows the crystal and the magnetic structures of \bnf .
Fig. 1(a) corresponds to the alignment of magnetic moments
parallel to the layers which is found in most FeAs-based compounds
\cite{7,7b,8}. One has to distinguish three orthogonal directions:
one is defined by the in-plane component of the magnetic
propagation vector (chosen here as (0.5,0.5,1)), [1,1,0] labelled
longitudinal in-plane, the second is [-1,1,0] transversal
in-plane, and the third [0,0,1] out-of-plane. Fig. 2 shows the
temperature dependence of nuclear and magnetic Bragg reflections
measured on \bnfdf . The magnetic peaks (0.5,0.5,1) and
(0.5,0.5,3) start to increase in intensity when cooling below the
first magnetic transition at 70\ K. There is a second magnetic
transition at lower temperature where both peaks further increase,
but this low-temperature intensity enhancement is much stronger
for the (0.5,0.5,1) Bragg peak. Already this different behavior of
intensities at the two transitions indicates a change in the
magnetic structure at the lower transition. Fig. 2 also shows the
temperature dependence of two nuclear Bragg peaks which both show
anomalies at the two magnetic transitions. This impact of the
magnetic ordering on the nuclear structure underlines close
coupling but clear interpretation of these structural effects is
not obvious. The emergence of orthorhombic domains in the
magnetically ordered phase has strong impact on extinction and
multiple diffraction conditions of nuclear Bragg reflections
\cite{13}. The up and down shifts of nuclear intensities at the
two transitions indicates that the structural changes appearing at
the high-temperature magnetic transition get suppressed at the
lower transition. The anomalies in the nuclear Bragg peaks are not
perfectly sharp as it can be expected for an extinction effect at
a structural transition, but the width of the anomaly amounts only
to a few degrees K so that any local variation of the Na doping
must be very limited.

\begin{figure}[t]
\begin{center}
\includegraphics*[width=.75\columnwidth]{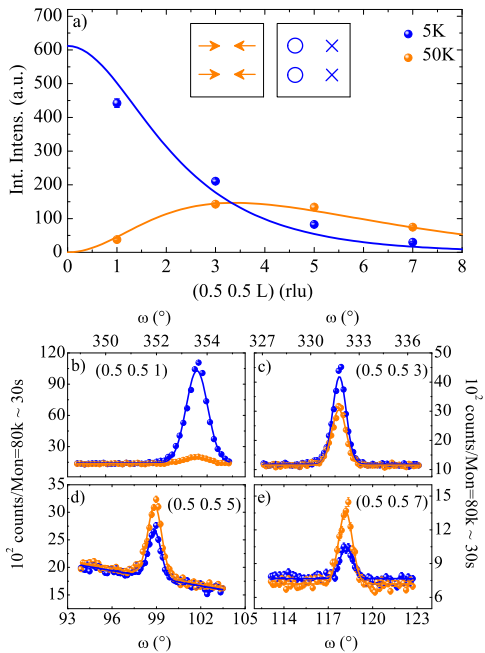}
\end{center}
\caption{(Color online) Part (a) shows the $l$ dependence of
magnetic integrated intensities in the AFM1 and AFM2 phases. The
lines are model calculations for the magnetic structures shown in
Fig. 1 with moments parallel to the $a$ and $c$ directions. (b-e)
Rocking scans across the magnetic Bragg peaks (0.5,0.5,$l$)  with
$l$=1,3,5,7, measured on the PANDA spectrometer at 5 and 50\ K, in
the AFM2 and AFM1 phases, respectively.   \label{fig3}}
\end{figure}

There is a third anomaly visible in the temperature dependencies
of the magnetic Bragg peaks. At low temperature in the
superconducting state both magnetic Bragg peaks decrease in
intensity, see Fig. 2.  The relative reduction in the two magnetic
peaks is identical indicating that this sample exhibits a single
magnetic phase at low temperature whose ordered moment gets
reduced in the superconducting state.

In order to quantitatively analyze the change in magnetic
structure we have measured integrated intensities on PANDA by
performing rocking scans. The integrated intensities were
corrected for the Lorentz factor and are traced in Fig. 3 against
the $q_l$ component. Neutron diffraction only measures the
magnetic moment perpendicular to the scattering vector. Any
magnetic diffraction intensity is weighted with a factor
$sin^2(\alpha)$ with $\alpha$ being the angle between the magnetic
moment and the scattering vector. By enhancing the $q_l$ component
in the magnetic Bragg reflection ${\bf Q}$=(0.5,0.5,$q_l$) a
magnetic moment aligned along the $c$ direction contributes less
and a moment along the in-plane component of the Bragg scattering
vector, i.e. along [110], contributes more strongly. In addition
the magnetic form factor depends on the length of the scattering
vector in an isotropic approximation and reduces the magnetic
scattering at larger $Q$ values. Both effects can be calculated
for ${\bf Q}$=(0.5,0.5,$q_l$) and are included in Fig. 3 for the
usual magnetic arrangement with moments aligned along the in-plane
component of the propagation vector (orange line, index $a$) and
for out-of-plane alignment (blue line, index $c$):

\begin{eqnarray}
I_{a,c}(0.5,0.5,q_l)\propto \left[ f(\vert Q\vert)\cdot m
\cdot sin(\alpha_{a,c}) \right]^2 \label{eq1} \\
sin(\alpha_{a})=\frac{q_l\cdot {\frac{2\pi}{c}}}{Q} \ ; \ 
sin(\alpha_{c})=\frac{{\frac{1}{\sqrt{2}}} \cdot
{\frac{2\pi}{a}}}{Q} \label{eq3}
\end{eqnarray}

At 50\ K, i.e. in the magnetic phase at higher temperature, the
integrated intensities are perfectly described by the in-plane
arrangement, while the low temperature data agree well with the
out-of-plane alignment. Some minor deviations may arise from an
anisotropic magnetic form factor or from a small fraction of the
sample staying in the AFM1 phase (fitting the data in Fig. 3 (a)
with both phases and equal moments indicates only 13\% AFM1).

We may conclude that \bnfdf \ exhibits a spin reorientation
transition at $\sim$46\ K with moments aligned along the $c$
direction at low temperature. The different intensity changes
observed at different (0.5,0.5,$q_l$) peaks are furthermore
illustrated in the lower parts of Fig. 3. The gain in intensity is
most pronounced at (0.5,0.5,1) where the geometry factor strongly
hampers the observation of the magnetic peaks in the AFM1 phase,
$sin^2(\alpha_a)$=0.16, while out-of-plane moments contribute with
an almost full geometry factor in AFM2, $sin^2(\alpha_c)$=0.84.
The scans across (0.5,0.5,5) and (0.5,0.5,7) show that these
intensities even decrease at the lower transition because the
geometry factor is strongly reduced at the spin reorientation.

\begin{figure}[t]
\begin{center}
\includegraphics*[width=.7\columnwidth]{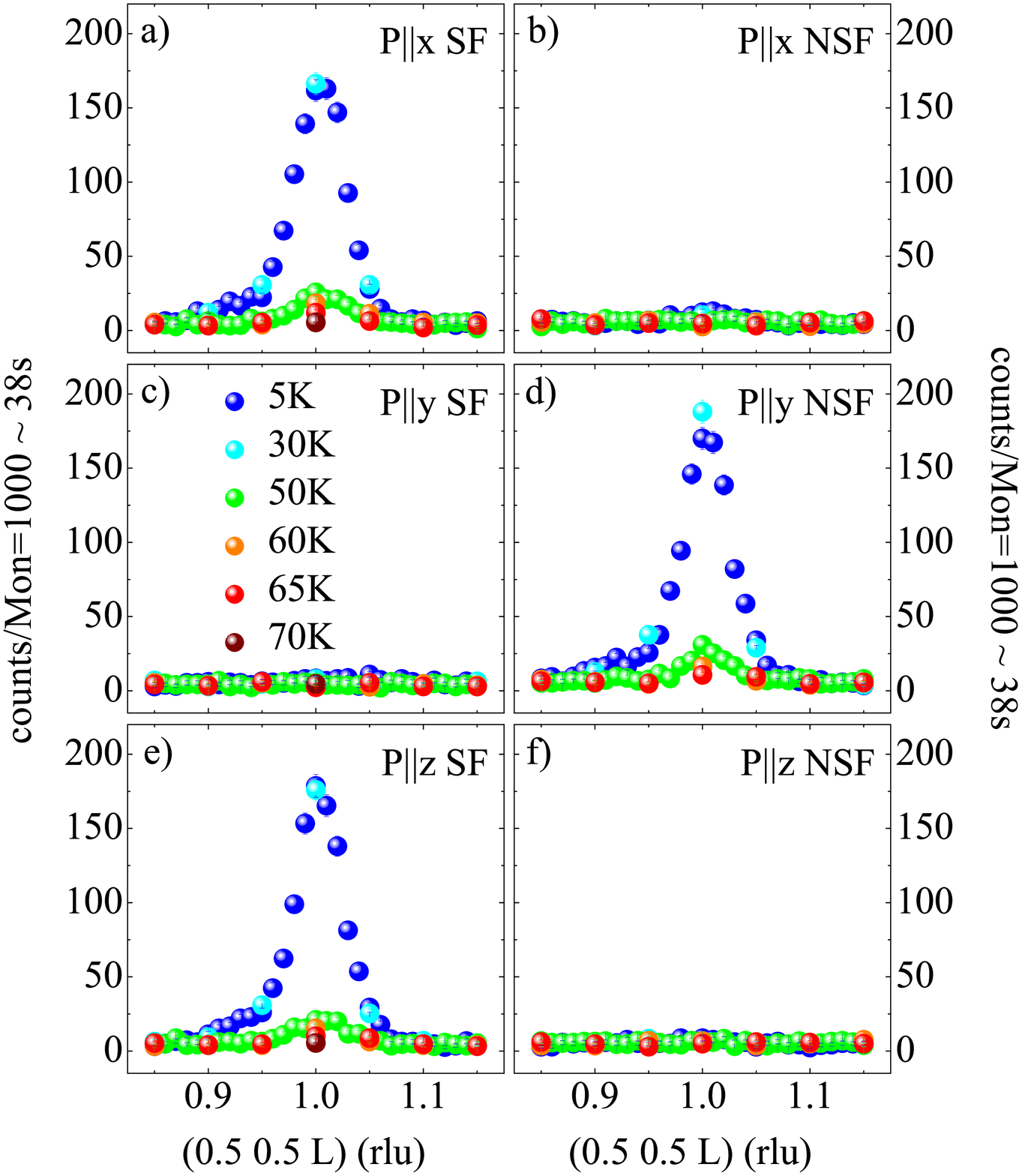}
\end{center}
\caption{ Results of  polarized neutron-diffraction experiments on
the 4F spectrometer. Scans were performed across the (0.5,0.5,1)
magnetic Bragg peak in both the SF and NSF channel for the neutron
polarization along the $x$, $y$ and $z$ directions. There is no
indication for any scattering associated with magnetic moment
pointing in the [1,-1,0] direction. \label{fig4}}
\end{figure}

We have strengthened our conclusion about the spin reorientation
in \bnfdf \ by experiments using polarized neutron scattering, see
Fig. 4. The sample was again mounted with the [110]/[001]
scattering plane. The polarization analysis adds the additional
selection rule that magnetic scattering contributes to the
spin-flip (SF) signal if the magnetic contribution is
perpendicular to the axis of the polarization analysis, while the
component parallel to the polarization axis contributes to the
non-spin-flip (NSF) scattering. One may analyze the SF and NSF
signals for setting the neutron-polarization axis parallel to the
scattering vector (labelled x-direction), parallel to the [$\bar
1$10] direction (labelled z-direction, as it is perpendicular to
the scattering plane) and perpendicular to these two directions
(labelled y-direction). Fig. 4 shows the measurement obtained with
the \bnfdf -crystal at the (0.5,0.5,1) reflection, measurements at
(0.5,0.5,3) yielded the same results. For x-axis polarization
analysis the scattering appears almost entirely in the SF channel,
as it is expected for a magnetic signal. The small leakage in the
NSF channel is due to imperfection of the polarization analysis,
but note that the flipping ratio of the total experimental setup
is quite high, $\sim$31. Concerning the other two polarization
directions the signal is entirely found in the NSF channel in
$y$-direction and in the SF channel in the $z$-direction. These
findings document that there is no ordered magnetic moment along
the [$\bar 1$10] direction (in-plane transversal) neither in the
high-temperature nor in the low-temperature phase. The magnetic
structure with the same propagation vector (0.5,0.5,1) but with
transversal in-plane moments is not observed in \bnfdf . The
magnetic moment thus always stays within the scattering plane.
Taking into account the $q_l$ dependence of the signal, the spin
reorientation from [110] to [001] directions is the only possible
explanation of our single-crystal data.

We have also studied another crystal with a 35\% Na doping as well
as ones with 25 and 39\%. All these crystals show qualitatively
the same two magnetic transitions and the spin reorientation.
However, the transition temperatures depend on the doping and at
25\% Na and 39\% Na contents the low temperature phase transition
seems to stay incomplete indicating the limits of the stability of
the $c$ aligned magnetic phase. A crystal with a doping content of
40\% did not show any static magnetic ordering indicating a rapid
suppression of the magnetic phase.

Avci et al. reported on powder neutron diffraction experiments on
a sample with nominal composition
Ba$_{0.76}$Na$_{0.24}$Fe$_2$As$_2$ \cite{14} which also indicate a
second magnetic transition in good agreement with our results.
This powder sample shows reentrance into an almost tetragonal
phase, but the powder data and the fact that 40\% of the sample
did not transform prohibited an unambiguous determination of the
magnetic structure. The results in \cite{14} were discussed as
evidence for a 2$k$ magnetic structure associated with the
simultaneous condensation of both nematic order parameters. Since
our results clearly document the spin reorientation as the main
element of the low-temperature magnetic transition the possible
existence of a 2$k$ structure needs further analysis, because such
structure implies non-magnetic or weakly magnetic sites for
$c$-aligned moments, and because the spin reorientation can at
least partially explain the suppression of the orthorhombic
splitting.

The spin reorientation observed in \bnf \ is in agreement with the
anisotropies found in pure and in doped \bfa \cite{12,13,13d}. The
analysis of the gaps of magnetic excitations in pure \bfa \
indicates that rotating the moment into the $c$ direction
corresponds to the lower anisotropy energy in contrast to the
expectation for a layered magnet. Na substitution seems to further
reduce the $c$-axis anisotropy energy so that it becomes the
ground state ordering for 35\% of doping. The change in the
direction of the magnetic moment must be associated with a
slightly different orbital arrangement.

The ARPES experiments in the AFM phase of \bfa \ indicate
significant lowering of the electronic band associated with
$d_{xz}$ orbitals (in the coordinates with $a$ along the AFM
ordered moment) \cite{9a}. In a simple localized picture, orbital
ordering with enhanced occupation of the $d_{xz}$ orbital can mix
through spin-orbit coupling with other orbitals yielding a finite
orbital moment. Linear combinations with the other $t_{2g}$
orbitals $d_{xz}+i\cdot d_{yz}$ and $d_{xz}+i\cdot d_{xy}$ yield
orbital moments along $z$ and $x$, respectively, which are the two
directions of ordered moments in AFM FeAs-materials. The enhanced
occupation of the $d_{xz}$ orbital observed in ARPES qualitatively
agrees with the observed ground states in the AFM phases but a
quantitative analysis of the spin-orbit coupling in the metallic
AFM state is highly desirable. Avci et al. \cite{16} find that the
FeAs layers are more extended in $c$ direction in the Na-doped
series, which favors the orientation of magnetic moments in $c$
direction giving further support to our orbital physics
argumentation.

The different magnetic ground state seems not to be reflected by a
quantitative change in the superconducting T$_c$ as compared to
the K-doped series \cite{15,16,17,bkfa}. This fact is a challenge
for any quantitative pairing theory in particular one based on
orbital fluctuations. The spin reorientation, furthermore, is
relevant for discussing the possible role of the nematic phase
\cite{nematic}. At least the breaking of the fourfold axis due to
the in-plane alignment of the ordered moment seems not to possess
a significant effect on the superconducting pairing. The two
different magnetic arrangements appearing in FeAs-compounds,
furthermore, indicate that the magnetic character in these
materials is not Ising like. Instead orbital effects seems just to
imply a magnetic hard axis (transversal in-plane).

The $c$-polarized phase seems, furthermore, to agree with the
recent observation of a sharp low-energy resonance mode in
optimally Co-doped \bfa , which arises in the $c$-polarized
channel \cite{13}. This experiment showed that this low-energy
mode does not appear in the in-plane transversal direction, while
a similar experiment on a Ni-doped \bfa \ sample \cite{13d}
indicates that the low-energy mode also has some longitudinal
in-plane character. Therefore, one may conclude that also in the
doped superconducting compounds the lowest magnetic excitations
are found in in-plane longitudinal and in c-oriented channels
(easy plane) while in-plane transversal excitations lie higher in
energy. The low-energy excitations in superconducting samples
correspond thus to the two directions, where ordered moments are
observed in the static phases.

So far only \bnf \ exhibits the c-polarized order, but the
high-pressure phase recently reported in K-doped \bfa \ could have
the same character \cite{hassinger}. Magnetic order in the
transversal in-plane direction is never observed and the
corresponding magnetic response in the superconducting phase lies
higher in energy underlining a common orbital effect. The
transversal in-plane direction remains the magnetic hard axis
independently of the presence of static magnetic order.

In conclusion we have shown that \bnf \ exhibits a spin
reorientation at low temperature. The magnetic order in \bnfdf \
corresponds to that observed in most other FeAs-based magnetic
materials for temperatures between 46 and 70\ K, but at 46\ K
magnetic moments reorient along the $c$ direction. This different
magnetic structure seems not to be reflected in a modified
superconducting pairing as superconducting transition temperatures
in Na and in K doped systems are comparable. AFM and
superconducting FeAs-based compounds exhibit evidence for the same
and sizeable easy-plane anisotropy underlining the importance to
take spin-orbit coupling into account.

This work was supported by the Deutsche Forschungsgemeinschaft
(DFG) through the Priority Programme SPP1458 (Grants No.
BE1749/13, BU887/15-1 and BR2211/1-1). S.W. thanks the DFG for
funding in the Emmy Noether Programme (project 595/3-1). We thank
D. Khomskii for valuable discussions about orbital aspects.


\end{document}